\documentclass[aps,prl,twocolumn,amsmath,amssymb,showpacs,superscriptaddress,notitlepage,longbibliography]{revtex4-1}
\pdfoutput=1
\usepackage[colorlinks=true,linkcolor=blue,anchorcolor=red,citecolor=blue, urlcolor=blue]{hyperref}
\usepackage{bm}
\usepackage{graphicx}
\usepackage{color}

\UseRawInputEncoding

\begin{document}

\title{ Role of hidden spin polarization in non-reciprocal transport of antiferromagnets}

\author{Weizhao Chen}
\affiliation{Shenzhen Institute for Quantum Science and Engineering and Department of Physics, Southern University of Science and Technology, Shenzhen 518055, China}
\author{Mingqiang Gu}
\affiliation{Shenzhen Institute for Quantum Science and Engineering and Department of Physics, Southern University of Science and Technology, Shenzhen 518055, China}
\author{Jiayu Li}
\affiliation{Shenzhen Institute for Quantum Science and Engineering and Department of Physics, Southern University of Science and Technology, Shenzhen 518055, China}
\author{Panshuo Wang}
\affiliation{Shenzhen Institute for Quantum Science and Engineering and Department of Physics, Southern University of Science and Technology, Shenzhen 518055, China}
\author{Qihang Liu}
\email{Corresponding Author: liuqh@sustech.edu.cn}
\affiliation{Shenzhen Institute for Quantum Science and Engineering and Department of Physics, Southern University of Science and Technology, Shenzhen 518055, China}
\affiliation{Guangdong Provincial Key Laboratory of Computational Science and Material Design, Southern University of Science and Technology, Shenzhen 518055, China}
\affiliation{Shenzhen Key Laboratory of Advanced Quantum Functional Materials and Devices, Southern University of Science and Technology, Shenzhen 518055, China}

\date{\today}

\begin{abstract}
The discovery of hidden spin polarization (HSP) in centrosymmetric nonmagnetic crystals, i.e., spatially distributed spin polarization originated from local symmetry breaking, has promised an expanded material pool for future spintronics. However, the measurements of such exotic effects have been limited to subtle space- and momentum-resolved techniques, unfortunately hindering their applications. Here, we theoretically predict macroscopic non-reciprocal transports induced by HSP when coupling another spatially distributed quantity, such as staggered local moments in a \textit{PT}-symmetric antiferromagnet. By using a four-band model Hamiltonian, we demonstrate that HSP plays a crucial role in determining the asymmetric bands with respect to opposite momenta. Such band asymmetry leads to non-reciprocal nonlinear conductivity, exemplified by tetragonal CuMnAs via first-principles calculations. We further provide the material design principles for large nonlinear conductivity, including two-dimensional nature, multiple band crossings near the Fermi level, and symmetry protected HSP. Our work not only reveals direct spintronic applications of HSP (such as N\'{e}el order detection), but also sheds light on finding observables of other ¡°hidden effects¡±, such as hidden optical polarization and hidden Berry curvature.
\end{abstract}

\maketitle

\emph{Introduction}.\rule[3pt]{0.4cm}{0.02em}According to the seminal works of Dresselhaus and Rashba, spin polarization in nonmagnetic crystals that lack inversion symmetry \textit{P} is possible due to the relativistic spin-orbit coupling (SOC) \cite{Ref.1,Ref.2}. Recent studies revealed that a related form of ¡°hidden spin polarization¡± (HSP) should exist even in centrosymmetric crystals, as long as atomic sites individually break inversion symmetry locally \cite{Ref.3,Ref.4}. Such an HSP effect could significantly expand the pool of potential materials for spintronics \cite{Ref.5,Ref.6}, and provide new physical insights into the relevant hidden physics such as optical polarization, valley polarization, orbital polarization, Berry curvature, etc \cite{Ref.7,Ref.8,Ref.9,Ref.10,Ref.11,Ref.12,Ref.13,Ref.14,Ref.15,Ref.16,Ref.17}. Owing to the coexistence of \textit{P} and time-reversal symmetry \textit{T}, the HSP in nonmagnetic materials localized at a sector $\bm{P}_{k}(r)$ is odd distributed in both real and momentum spaces, i.e., $\bm{P}_{k}(r)=-\bm{P}_{k}(-r)$, and $\bm{P}_{k}(r)=-\bm{P}_{-k}(r)$. Therefore, while HSP has been successfully measured by techniques with both $\bm{r}$- and $\bm{k}$-resolution, such as spin- and angle-resolved photoemission spectroscopy (spin-ARPES) \cite{Ref.18,Ref.19,Ref.20,Ref.21,Ref.22,Ref.23,Ref.24,Ref.25,Ref.26}, its application in spintronics requires global symmetry breaking (e.g., via external electric fields) to release the compensation of polarization originated from the high global symmetry \cite{Ref.4,Ref.Tao2019}.

To utilize the HSP effect for a macroscopic observation requires breaking both sector and momentum compensation conditions described above, rendering a challenging issue if the global symmetry is maintained. Here we propose a way to manipulate HSP $\bm{P}_{k}(r)$ by coupling another odd-distributed quantity $\bm{M}(r)$. The resultant observable $O(r)=\int\bm{P}_{k}(r)\cdot\bm{M}(r)$ yields a superposition effect between the two sectors rather than compensation. Specifically, when $\bm{M}(r)$ represents the staggered N\'{e}el order in antiferromagnetic (AFM) materials, the absence of \textit{T} also removes the momentum compensation, leading to possible spin-orbitronic applications. Recently, the coupling between the AFM order and SOC-induced spin texture is predicted to induce a spin-orbit torque that can in turn switch the N\'{e}el order electrically in antiferromagnets CuMnAs and Mn$_{2}$Au \cite{Ref.27,Ref.28}. However, a following experimental study suggested that the thermal effect may reduce the reliability of such electrical switching \cite{Ref.29}.

In this work, we investigate the non-reciprocal transport in \textit{PT}-symmetric antiferromagnets, in which every band is spin degenerate, yet exhibiting HSP effects due to the local symmetry breaking. Starting from a four-band model and Boltzmann transport theory, we elucidate that by coupling the staggered AFM magnetization, HSP is not only the trigger of non-reciprocal transport but also determines the magnitude of the non-reciprocal nonlinear current (NLC) via the pivot of band asymmetry between opposite momenta $\pm \bm{k}$. To demonstrate our theory, we apply first-principles calculations to simulate the NLC of tetragonal CuMnAs, yielding qualitatively consistent results with the recent experiment \cite{Ref.32}, and quantitatively identifying the crucial role of HSP. We then provide the design principle of antiferromagnets with large NLC, including large HSP that could be protected by nonsymmorphic symmetry, band crossings/anticrossings around the Fermi surface, and the nature of low dimension. Our finding reveals a macroscopic observable of HSP whose identification is previously limited to space and momentum resolved measurements, paving an avenue for HSP materials in the applications of future AFM spintronics.

\begin{figure}[thb!]
\centering
\includegraphics[width=0.45\textwidth]{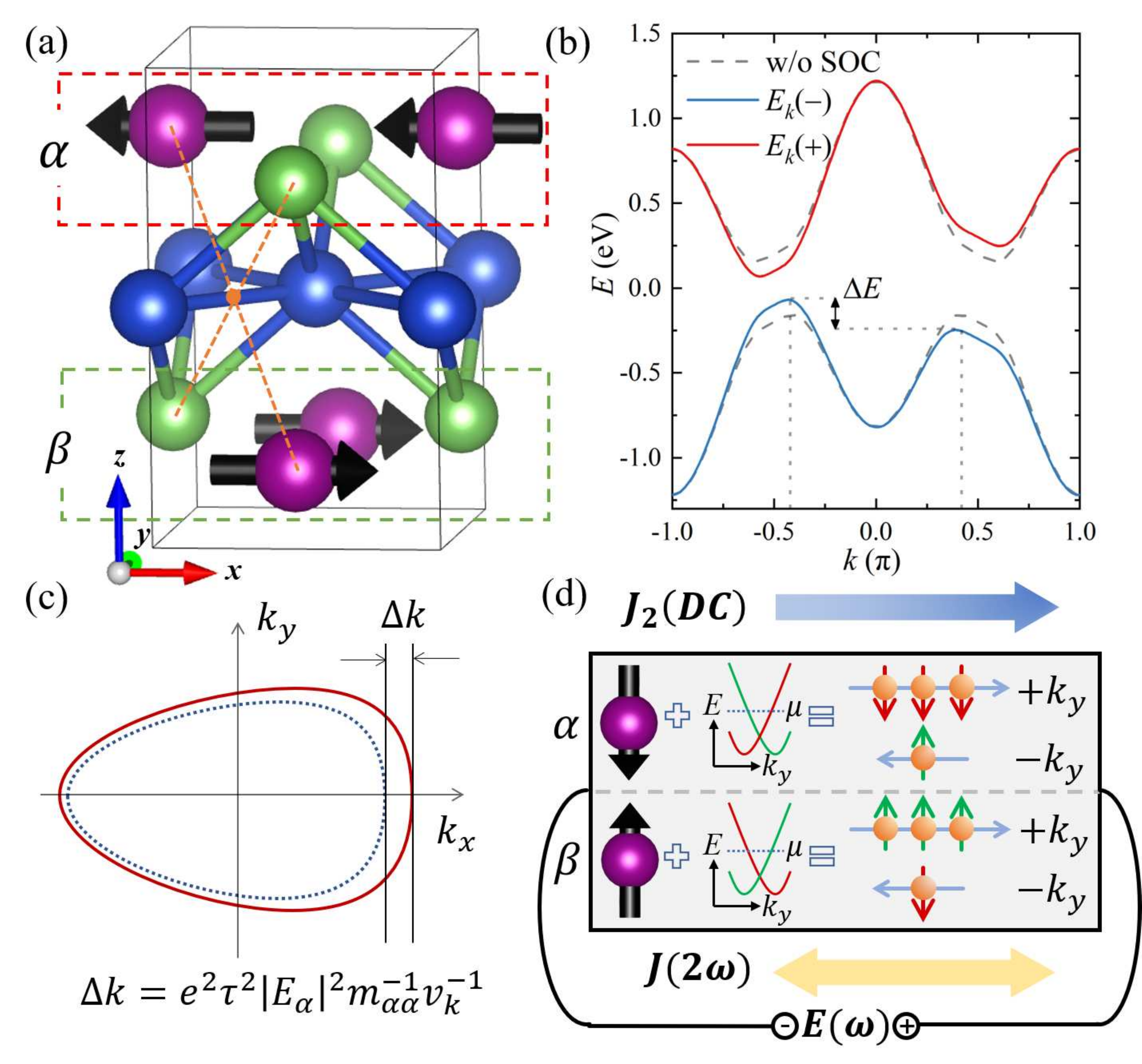}%
\caption{\label{fig1} \raggedright (a) Crystal structure of CuMnAs as an example of \textit{PT}-symmetric antiferromagnets with two sectors $\alpha$ and $\beta$ connected by \textit{PT} center with orange dashed lines. (b) The band structure of the four-band Hamiltonian with band asymmetry. The parameters used for calculation are $t_k=\cos^2 k$, $\epsilon_k=0.2\cos k$, $M=0.2$, and $\lambda_k=0.1\sin k$ for SOC. (c) Schematic plot of the Fermi surface expansion/shrinking due to the second-order effect of the electric field $E_\alpha$. (d) Schematic plot of the formation of longitudinal nonlinear current, which is originated from the coupling between HSP and staggered magnetization.  The red bands possess spin down polarization as the red arrows adhere to the carriers on the right hand, and spin up polarization is denoted by the green.}
\end{figure}

\emph{Coupling between hidden SOC and N\'{e}el order}.\rule[3pt]{0.4cm}{0.02em}To elucidate the coupling between the HSP effect and N\'{e}el order, we consider \textit{PT}-symmetric antiferromagnets. While the global \textit{PT} symmetry ensures spin degeneracy throughout the whole Brillouin zone, in real space there could be ``sectors'' breaking \textit{PT} locally and thus manifesting spin polarization owing to the relativistic SOC anchored on the nuclear sites. Such a concept of HSP is originally proposed and classified via crystallographic group in nonmagnetic materials with inversion symmetry \cite{Ref.3}, but can be easily extended to \textit{PT}-symmetric antiferromagnets. Taking the tetragonal structure of CuMnAs as an example [Fig. \ref{fig1}(a)], the two Mn atoms (site symmetry 4mm) in a unit cell with opposite in-plane magnetizations are symmetric with respect to the \textit{PT} center, thus belonging to two sectors $\alpha$ and $\beta$, respectively. Therefore, the HSP is defined by projecting the Bloch wavefunctions $\psi_{nk}$ of the doubly degenerate bands onto a specific sector, say $\alpha$, $\bm{P}_k^\alpha=\sum\limits_{i\in\alpha}\sum\limits_{n\in\mathbb{N}}\langle\psi_{nk}|\bm\sigma\bigotimes|i\rangle\langle i|\psi_{nk}\rangle$ , where $\bm{k}$, $\bm\sigma$, and $|i\rangle$ denote the wavevector, Pauli operator for spin degree of freedom, and the localized orbitals belonging to sector $\alpha$, respectively. $\mathbb{N}$ denotes the summation over the two degenerate bands, implying that $\bm{P}_k^{\alpha}$ is independent of the choice of the wavefunction basis.

We then construct a four-band model Hamiltonian containing both HSP and N\'{e}el order as follows:
\begin{widetext}
\begin{equation}
\begin{aligned}
  H = \begin{pmatrix}
      \epsilon_k^\alpha & ((\bm\lambda_k + \bm M)\cdot\ \bm \sigma)_{12} & t_k & 0 \\
      ((\bm\lambda_k + \bm M)\cdot\ \bm \sigma)_{12}^\ast & \epsilon_k^\alpha & 0 & t_k \\
      t_k^\ast & 0 & \epsilon_k^\beta & -((\bm\lambda_k + \bm M)\cdot\ \bm \sigma)_{12} \\
      0 & t_k^\ast & -((\bm\lambda_k + \bm M)\cdot\ \bm \sigma)_{12}^\ast & \epsilon_k^\beta \\
      \end{pmatrix},
\end{aligned}
\end{equation}
\end{widetext}
where $\epsilon_k^\alpha=\epsilon_k^\beta=\epsilon_k$ is the on-site energy and $t_k$ is the wavefunction hybridization between the two sectors. $\bm\lambda_k$ and $\bm M$ denote the effective magnetic field caused by SOC and the local moment, respectively. For example, classic Rashba and Dresselhaus effects correspond to $\bm\lambda_k=\alpha_R (k_y,-k_x,0)$ and $\bm\lambda_k=\alpha_D (k_x,-k_y,0)$, respectively. The subscript `12' represents the off-diagonal element of the 2-rank matrix, i.e., $((\bm\lambda_k + \bm M)\cdot\ \bm \sigma)_{12}= (\bm\lambda_{k,x} + \bm M _{x}) - i(\bm\lambda_{k,y} + \bm M _{y})$. Since \textit{PT}-symmetry is conserved, there are the two sets of doubly degenerate bands with the eigenvalues $E_k (\pm)=\epsilon_k\pm\sqrt{|\bm\lambda_k + \bm M|^2+|t_k|^2}$, as shown in Fig. \ref{fig1}(b). The corresponding HSP are $\bm P_k^¦Á (\pm)=\pm(\bm\lambda_k + \bm M)/\sqrt{|\bm\lambda_k + \bm M|^2+|t_k|^2}$, indicating that the hybridization between sectors $t_k$ is detrimental to the magnitude of HSP.

A key condition to have non-zero macroscopic observable such as non-reciprocal carrier transport is the band asymmetry with respect to opposite vectors, i.e., $E_k \neq E_{-k}$ \cite{Ref.31,Ref.33}. While such condition is fulfilled when both \textit{P} and \textit{T} are broken, its magnitude, namely, $\Delta E=E_k-E_{-k}$, is not determined by symmetry arguments and remains unclear. According to Eq. (1), we analytically obtain:
\begin{equation}
\begin{aligned}
  \Delta E (\pm) = \frac{\pm 4 P_k^{ave} \bm\lambda_k \cdot \bm M}{|\bm\lambda_k + \bm M|+|\bm\lambda_k - \bm M|},
\end{aligned}
\end{equation}
where $P_k^{ave}=\{w|P_{+k}^\alpha |+(1-w)|P_{-k}^\alpha |\}$ is the weighted average of HSP with the weight $w=\sqrt{|\bm\lambda_k + \bm M|^2+|t_k|^2}/\{\sqrt{|\bm\lambda_k + \bm M|^2+|t_k|^2} +\sqrt{|\bm\lambda_k - \bm M|^2+|t_k|^2}\}$. From Eq. (2) we find that when $\bm\lambda_k$ is parallel to $\bm M$, the band asymmetry $\Delta E=2min\{|\bm\lambda_k|,|\bm M|\}P_k^{ave}$ is proportional to the HSP. On the other hand, $\Delta E$ vanishes when $\bm\lambda_k$ is perpendicular to $\bm M$. Eq. (2) clearly shows that in \textit{PT}-symmetric antiferromagnets, the cooperation of magnetic order, SOC and local symmetry breaking as the key reason gives rise to band asymmetry, which further leads to a global non-reciprocal transport behavior, as discussed in the following.

\emph{Non-reciprocal nonlinear conductivity}.\rule[3pt]{0.4cm}{0.02em}Unlike the first-order current caused by the drifting move, the second-order effect of the electric field $\bm E$ originates from the expansion or shrinking of the Fermi surface due to carrier acceleration. In this way, an asymmetric Fermi surface caused by band asymmetry leads to a net nonlinear current [Fig. \ref{fig1}(c)]. Expanding the Boltzmann equation to the second-order with respect to $\bm E$ (see Supplementary Section A for details \cite{Ref.supplementary}), we obtain a nonlinear current $j_{2,\gamma}=\sum\limits_{\alpha\beta}\sigma_{\gamma\alpha\beta} E_\alpha E_\beta$ with the second-order conductivity tensors for an energy band as follows:
\begin{equation}
\begin{aligned}
  \sigma_{\gamma\alpha\beta} = \frac{ e^3 \tau^2}{ V N_k } \sum\limits_{k \in BZ} m_{\gamma\alpha}^{-1} v_{\beta} \frac{\partial f_0}{\partial \epsilon},
\end{aligned}
\end{equation}
where $\tau$ is the relaxation time and $m_{\gamma\alpha}^{-1}=\frac{1}{\hbar} \frac{\partial v_\gamma}{\partial k_\alpha}$ is the inverse effective mass with the group velocity $v$. We note that Eq. (3) describes the extrinsic NLC, which is equivalent to the other forms of the Drude term used in recent studies \cite{Ref.34,Ref.35,Ref.36,Ref.Fu2015}. Nevertheless, Eq. (3) explicitly highlights the effective mass as another key ingredient of the nonlinear current.

As shown in Fig. \ref{fig1}(d), given a nonzero $\sigma_{\gamma\alpha\beta}$, an electric field $\bm E(\omega)$ would drive a nonlinear current $j_{2,\gamma}$ composed of an alternating current $J(2\omega)$ with the frequency of 2$\omega$ and a direct current $J_2$, leading to the unidirectional magnetoresistance effect \cite{Ref.37,Ref.38,Ref.39}. Specifically, for CuMnAs, the forward and backward currents in the $y$-direction carry opposite spin polarization ($\pm x$) due to the Rashba-type spin texture and then are scattered by the atomic magnetic moments (along $x$), resulting in a unidirectional current in a single sector. For the other sector, the cooperation of both reversed HSP and magnetic moment leads to a coherent effect. As a result, a longitudinal nonlinear current emerges along the $y$ direction. With the symmetry analysis of Eq. (3) (see Supplementary Section B \cite{Ref.supplementary}), nonzero $\sigma_{yxx}$ and $\sigma_{yzz}$ for the tetragonal CuMnAs are also allowed, indicating that both $E_x$ and $E_z$ can drive Hall-type nonlinear currents along $y$. The nonlinear current has been proposed to be a rectifier \cite{Ref.30,Ref.31,Ref.33}, while in antiferromagnets it can also be used as the reading head of the N\'{e}el order \cite{Ref.32,Ref.34,Ref.40}. For a 180$^\circ$ rotation of N\'{e}el order, the local moments $\bm M$ are reversed while the local SOC fields $\bm\lambda_k$ are conserved. As a result, the coupling-induced NLC changes its sign. We note that such HSP-assisted NLC exists in antiferromagnets with \textit{PT}-symmetry, which forbids the nonlinear anomalous Hall effect originated from Berry curvature dipole \cite{Ref.40}.

To further investigate the relationship between band asymmetry and nonlinear current, we simplify Eq. (3) and consider $\sigma_{yyy}$ by assuming uniform dispersions along $k_x$ and $k_z$,

\begin{equation}
\begin{aligned}
  \sigma_{yyy} &= \frac{ e^3 \tau^2}{ 2 \pi V } \{m_{yy}^{-1}(k_F^{-})-m_{yy}^{-1}(k_F^{+})\}\\
   &\approx \frac{ e^3 \tau^2}{ 2 \pi V } \{ \frac{\partial m_{yy}^{-1}}{\partial \epsilon} (k_F^{-})\Delta E + \frac{\partial^2 \Delta E}{\partial k ^2} \},
\end{aligned}
\end{equation}
where $k_F^{+} (k_F^{-})$ is the Fermi vector with the positive (negative) group velocity. Eq. (4) directly explains the relationship between NLC and the band asymmetry $\Delta E=E(k_F^{-})-E(-k_F^{-})$, which is characterized by HSP as discussed above. In addition, the first term indicates that large changes in effective mass, such as the tipping point caused by band crossing or anticrossing, could also enhance the conductivity. We next perform  NLC calculations based on atomistic Hamiltonians obtained from density-functional theory (DFT) to further illustrate the role of HSP in non-reciprocal transports (see Supplementary Section B \cite{Ref.supplementary}).

\emph{Nonlinear conductivity in CuMnAs}.\rule[3pt]{0.4cm}{0.02em}We choose CuMnAs to identify the characteristics of nonreciprocal NLC and the role of HSP because it is a room-temperature, \textit{PT}-symmetric, antiferromagnetic metal with intriguing Dirac fermions in various structures \cite{Ref.41,Ref.CuMnAs2017,Ref.42}, rendering a nice platform for novel transport properties. While recent works have predicted its intrinsic nonlinear Hall effect, which is originated from the Berry connection \cite{Ref.43,Ref.44}, we focus on the extrinsic current related to impurities scattering that can be either longitudinal or transverse. In the following, we focus on the discussion of longitudinal NLC, i.e., $\sigma_{yyy}$, and leave the transverse NLC component $\sigma_{yxx}$ to Supplementary Section C \cite{Ref.supplementary}.

\begin{figure}[thb!]
\centering
\includegraphics[width=0.5\textwidth]{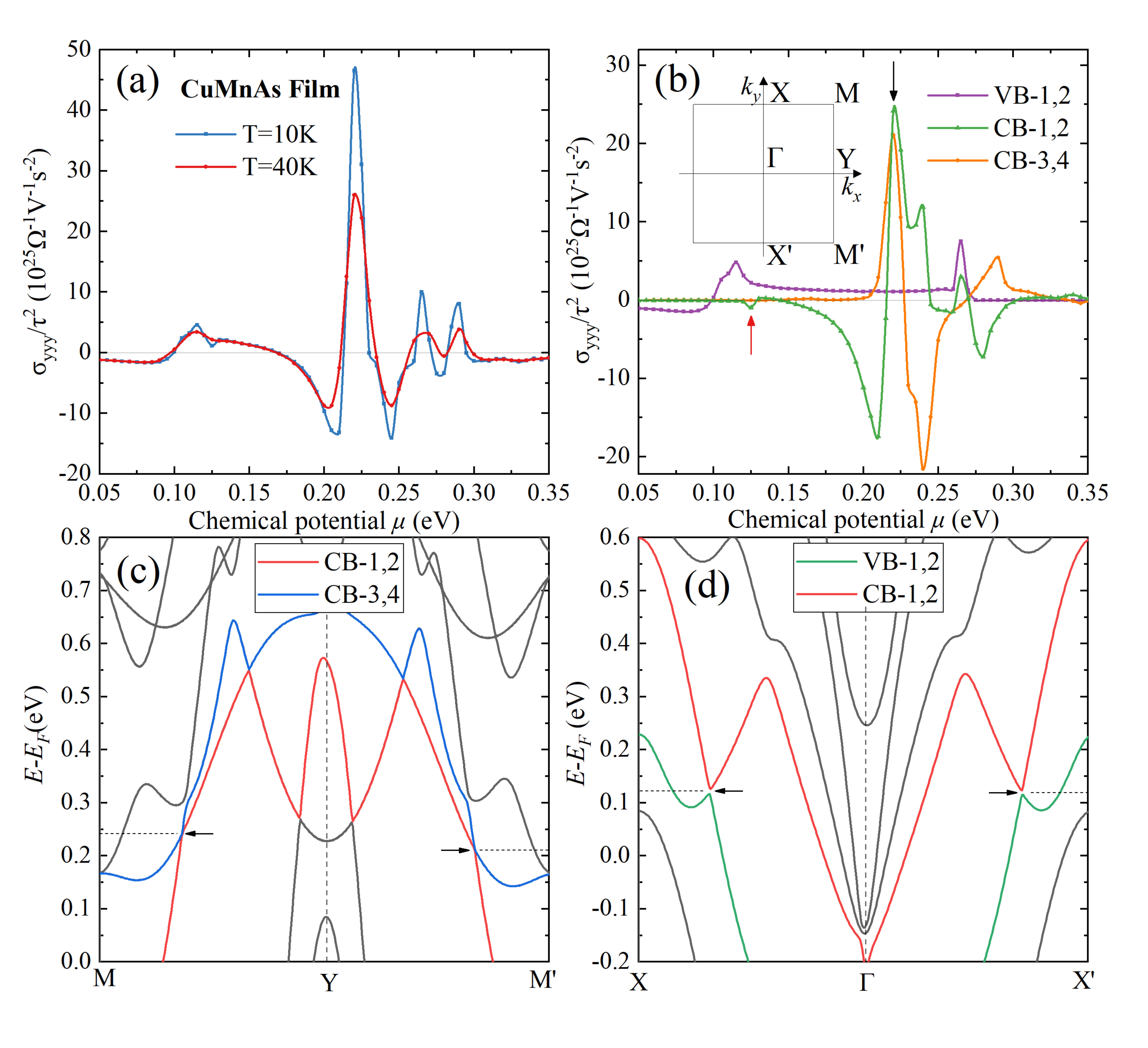}%
\caption{\label{fig2} \raggedright (a) Nonlinear conductivity $\sigma_{yyy}$ of CuMnAs film with a thickness of two primitive cells as a function of the chemical potential $\mu$ with different temperatures. (b) The band-decomposed $\sigma_{yyy}$ of CuMnAs film, where a small peak and a large peak for CB-1,2 are marked by the red and black arrows, respectively. The inset shows the two-dimensional Brillouin zone. (c) Band structure for the $k$-path along M-M', where the type-\uppercase \expandafter {\romannumeral 2} Dirac points are marked by the black arrows. (d) Band structure for the $k$-path along X-X', where the band anticrossings are marked by the black arrows.}
\end{figure}

We first calculate the NLC in a CuMnAs thin film with the thickness of two primitive cells and discuss the results of the bulk later. The calculated $\sigma_{yyy}$ as a function of the chemical potential $\mu$ is shown in Fig. \ref{fig2}(a), where a large and sharp peak appears at $\mu = 0.22$ eV accompanied by some oscillations. To reveal the origin of this peak, we decompose $\sigma_{yyy}$ onto the contributions from different bands around the Fermi energy. Three pairs of doubly degenerate bands, i.e., the highest valence bands VB-1,2, and the lowest conduction bands CB-1,2 and CB-3,4 are thus involved. As shown in Fig. \ref{fig2}(b), around $\mu = 0.22$ eV the NLC from CB-1,2 shows a negative peak followed by a positive one, while that of CB-3,4 shows positive and negative peaks sequentially. The two positive peaks happen to overlap each other and contribute to the peak of total NLC as denoted by the black arrow in Fig. \ref{fig2}(b).

From the band structure shown in Fig. \ref{fig2}(c), we find that the band asymmetry between the M-Y and M'-Y ensures the existence of the NLC. Specifically, the Dirac cone along M-Y is 0.029 eV higher than that along M'-Y, as denoted by the black arrows in Fig. \ref{fig2}(c). These type-\uppercase \expandafter {\romannumeral 2} Dirac points, formed by CB-1,2 and CB-3,4 around $\mu = 0.22$ eV, lead to the large NLC peaks in Fig. \ref{fig2}(b). Furthermore, the discontinuous effective masses induced by the Dirac points significantly enhance the NLC effect according to Eq. (4). We also notice that there is another small peak at $\mu \sim 0.12$ eV for the decomposed NLC from CB-1,2, as denoted by the red arrow in Fig. \ref{fig2}(b). This peak is caused by the band anticrossing between VB-1,2 and CB-1,2 along the $k$-path X-X' as shown in Fig. \ref{fig2}(d).

\begin{figure}[thb!]
\centering
\includegraphics[width=0.5\textwidth]{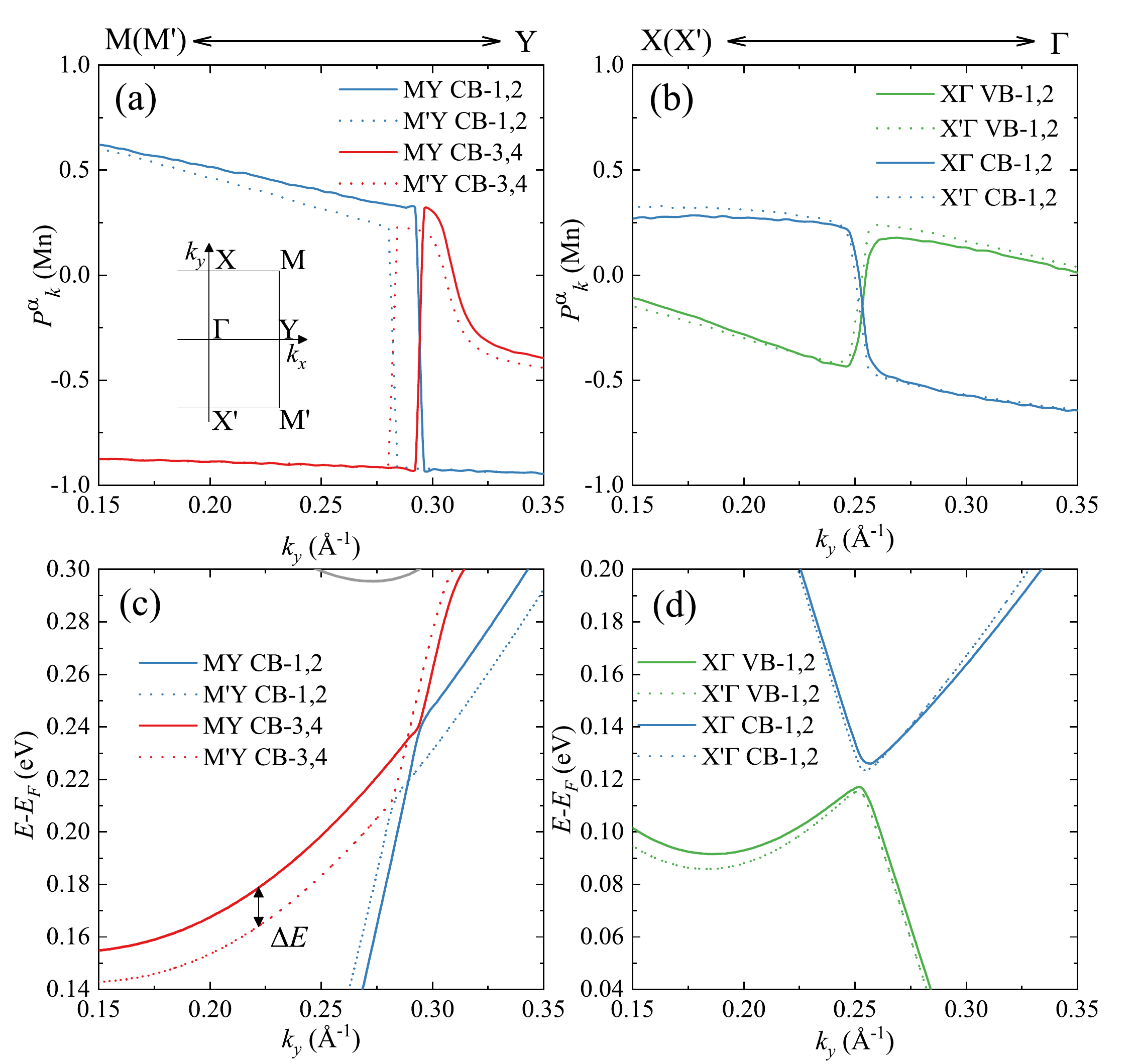}%
\caption{\label{fig3} \raggedright (a,b) The magnitude of HSP $\bm P_k^\alpha$(Mn) for the corresponding bands in the $k$-path along (a) M-M' with the type-\uppercase \expandafter {\romannumeral 2} Dirac points and (b) X-X' with the band anticrossings. (c,d) The band structures of CuMnAs film for the $k$-path along (c) M-M' with the type-\uppercase \expandafter {\romannumeral 2} Dirac points and (d) X-X' with the band anticrossings. The bands are folded with respect to Y and $\Gamma$ to show the band asymmetry $\Delta E$ clearly. The solid and dashed lines represent the $k$-path M-Y (X-$\Gamma$) and M'-Y (X'-$\Gamma$), respectively.}
\end{figure}

We next project the Bloch wavefunctions onto the magnetic Mn atoms and calculate the HSP $\bm P_k^\alpha$(Mn). Fig. \ref{fig3}(a) shows $\bm P_k^\alpha$(Mn) of CB-1,2 and CB-3,4 along M-Y-M', where two Dirac points are clearly presented with a momentum offset, indicating the band asymmetry. We find that both bands manifest large HSP, with one of them approaching almost 100\% polarization. In sharp contrast, $\bm P_k^\alpha$(Mn) of VB-1,2 and CB-1,2 along X-$\Gamma$-X' is much smaller, as shown in Fig. \ref{fig3}(b). Such significant difference in HSP is caused by the nonsymmorphic structure of CuMnAs. The glide mirror reflection minimizes the spin compensation between different sectors at the boundary of the Brillouin zone M-Y-M', thus retaining the HSP in each sector \cite{Ref.6,Ref.26}.

As shown in Eq. (2), the HSP closely correlates the band asymmetry $\Delta E$. This is further illustrated in Figs. \ref{fig3}(c) and \ref{fig3}(d), where we fold the band structures along M-Y-M' and X-$\Gamma$-X' with respect to Y and $\Gamma$, respectively, with the bands with $\pm k_y$ plotted on top of each other. As shown in Fig. \ref{fig3}(c), the large band asymmetry $\Delta E$ (about 17 meV) is attributed to the nearly 100\% $\bm P_k^\alpha$(Mn) around the Dirac cone, while $\Delta E$ around the band anticrossing in Fig. \ref{fig3}(d) is rather small with only 4 meV. As a result, the comparison between $\Delta E$ gives rise to the sharp contrast of the two peaks of NLC in Fig. \ref{fig2}(b), fully consistent with Eq. (4). The above analysis clearly highlights the significant role of HSP in the generation of nonlinear current: The required local symmetry breaking determines the existence of the nonlinear current in the AFM system with \textit{PT} symmetry; more importantly, it is also a key factor to the magnitude of NLC. Since the NLC is an effect regarding the Fermi surface, the critical temperature depends on the competition between $\Delta E$ and the average thermal energy of carriers. Therefore, HSP also has an effect that resists the decreasing of NLC upon increasing temperature (see Supplementary Section D \cite{Ref.supplementary}).

\begin{figure}[thb!]
\centering
\includegraphics[width=0.5\textwidth]{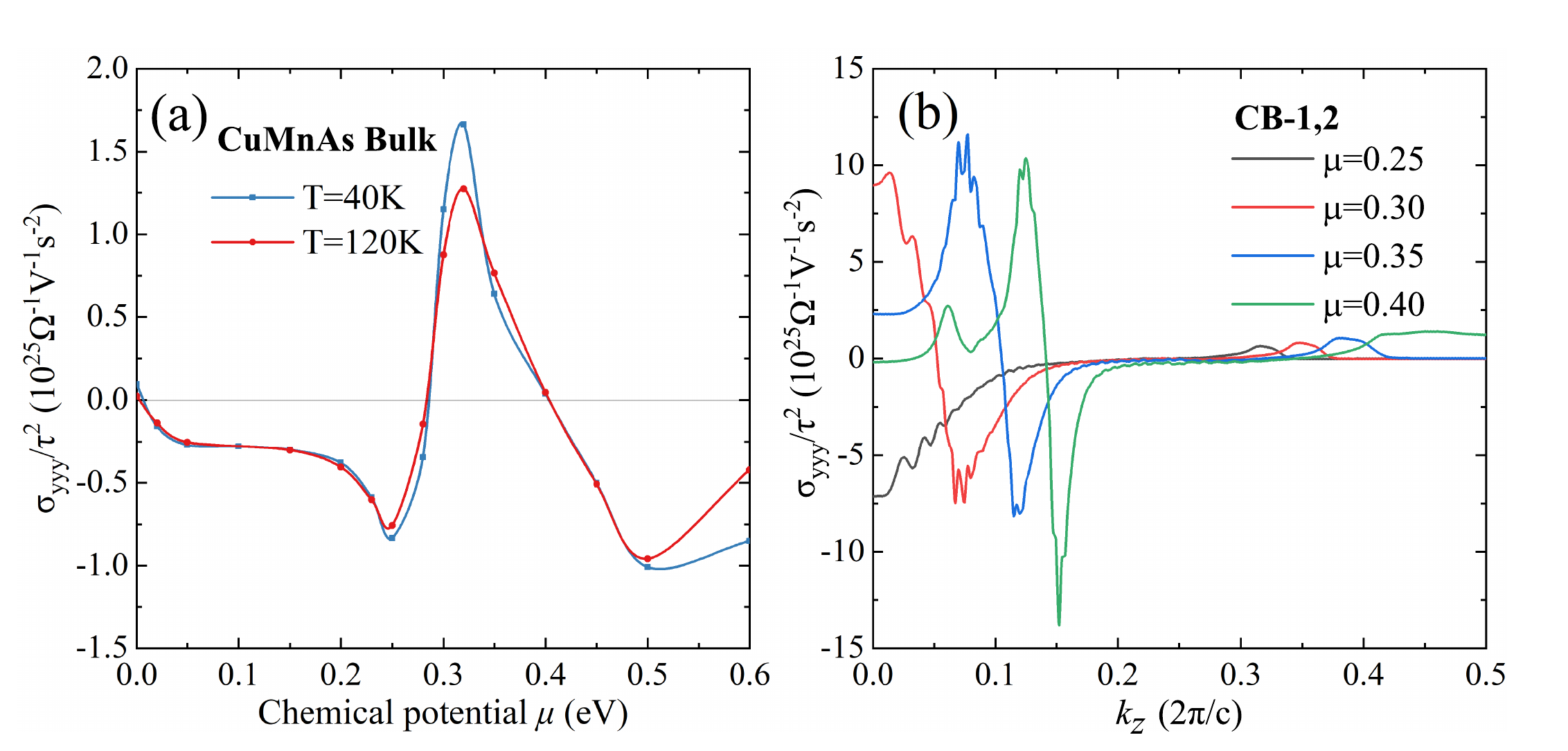}%
\caption{\label{fig4} \raggedright (a) Nonlinear conductivity $\sigma_{yyy}$ of Bulk CuMnAs as a function of chemical potential $\mu$ with different temperatures. (b) The $k_z$-resolved conductivity $\sigma_{yyy}$ with different chemical potential $\mu$.}
\end{figure}

\emph{Design principles}.\rule[3pt]{0.4cm}{0.02em}Before uncovering the design principles of large non-reciprocal NLC, we calculate the NLC of 3D bulk CuMnAs for comparison. Fig. \ref{fig4}(a) shows $\sigma_{yyy}$ as a function of $\mu$. It is found that similar with the 2D case, at around $\mu = 0.3$ eV the type-\uppercase \expandafter {\romannumeral 2} Dirac points induce a peak, which is one order of magnitude smaller than that of the thin film. To understand this, we resort to the $k_z$-resolved NLC. As shown in Fig. \ref{fig4}(b), the resolved $\sigma_{yyy}$ as the function of $k_z$ is composed of a positive peak followed by a negative peak, resulting in significant compensation over the integration. This feature essentially originates from the shift of Dirac cones in the energy scale with increasing $k_z$ (see Supplementary Section E \cite{Ref.supplementary}), implying that a small energy dispersion along $k_z$ or reducing the size from the bulk to thin film would suppress the compensation and thus enhance the magnitude of NLC.

Except for reducing the dimension, more design principles for \textit{PT}-symmetric antiferromagnets with advanced non-reciprocal transport properties are also revealed by the case of CuMnAs. Firstly, the NLC driven by HSP exists in all \textit{PT}-symmetric antiferromagnets with breaking \textit{P} in magnetic space groups (for 3D system) and breaking both \textit{P} and $C_{2z}$ in magnetic layer groups (for quasi-2D system). Secondly, nonsymmorphic symmetry could retain large HSP, which also determines the magnitude of band asymmetry and NLC. Moreover, the band crossing and anticrossing around the Fermi level serve as another amplifier of extrinsic NLC because of the large effective mass change. The abovementioned design principles would facilitate the search of material candidates with large NLC. Another candidate Mn$_2$Bi$_2$Te$_5$ is discussed in Supplementary Section F \cite{Ref.supplementary}.

\emph{Discussion}.\rule[3pt]{0.4cm}{0.02em}By coupling HSP with the N\'{e}el order of \textit{PT}-symmetric antiferromagnets, we show that a macroscopic measurable quantity, such as non-reciprocal transport, could emerge. Such finding expands the conventional detection of HSP that requires both space and momentum resolution, and proposes specific applications in antiferromagnetic spintronics, such as the identification and manipulation of N\'{e}el order \cite{Ref.32}. Experimentally, the N\'{e}el order of HSP antiferromagnets could be detected through either DC or AC measurement. The nonlinear longitudinal DC could be measured by the unidirectional magnetoresistance based on a double Hall cross with a four-probe configuration \cite{Ref.37}, yet without applying an external magnetic field; while the nonlinear AC for both longitudinal and transversal components could be measured based on a Hall cross structure with lock-in amplifiers \cite{Ref.32}.

More importantly, the basic idea of cooperating two spatially odd-distributed quantities opens a new avenue for exotic phenomena in the systems with various hidden polarizations. For example, coupling between the electron velocity (odd in position space) and the local spin polarization leads to the second-order spin photogalvanic effect and inverse Faraday effect \cite{Ref.45,Ref.46,Ref.47}, which can be understood as an optical response in the HSP system. A more generic picture can be grasped by considering the orbital Hall effect in centrosymmetric materials, which originates from the coupling of hidden Berry curvature and hidden orbital polarization \cite{Ref.48,Ref.49,Ref.Cysne2022}. Very recently, the nontrivial transport response originated from the coupling between the hidden Berry curvature and the N\'{e}el order is reported, dubbed as layer Hall effect \cite{Ref.50,Ref.Gu2022}. Overall, the non-reciprocal transport proposed here is a comprehensive demonstration of such effects, hopefully triggering future studies to unveil practical applications of hidden physical quantities.

This work was supported by National Key R\&D Program of China under Grant No. 2020YFA0308900, the National Natural Science Foundation of China under Grant Nos. 11874195 and 12274194, Guangdong Provincial Key Laboratory for Computational Science and Material Design under Grant No. 2019B030301001, the Science, Technology and Innovation Commission of Shenzhen Municipality (No. ZDSYS20190902092905285) and Center for Computational Science and Engineering of Southern University of Science and Technology.

\end{document}